\documentstyle[12pt]{article}
\begin{document}

\makebox[14cm][r]{TTP 93-39}\par
\makebox[14cm][r]{PITHA 93/44} \par
\makebox[14cm][r]{December 1993}\par
\vspace{.7cm}
\centerline{\Large \bf $K^0 $ Decay into Three Photons
\footnote{\it in celebration of the S.N. Bose birth centenary}}
\vspace{1.cm}
\centerline{P. Heiliger$^{a)} $, B. McKellar$^{b)} $  and L.M. Sehgal$^{c)} $}
\par
\centerline{$^{a)} $Institut f\"ur Theoretische Teilchenphysik} \par
\centerline{Universit\"at Karlsruhe}\par
\centerline{D--76128 Karlsruhe, Germany}\par
\centerline{$^{b)} $Research Center for High Energy Physics, School of
Physics,} \par
\centerline{University of Melbourne, Parkville, Victoria 3052, Australia} \par
\centerline{$^{c)} $III. Physikalisches Institut (A), RWTH Aachen}\par
\centerline{D--52056 Aachen, Germany}\par
\normalsize
\vspace{1.cm}

\begin{abstract}
The decays $K_{L,S} \to 3 \gamma $ are not forbidden by any selection rules or
symmetry principles. However, gauge invariance and Bose statistics dictate
that every photon pair in these transitions has at least two units of angular
momentum. This gives rise to an extraordinary suppression. Using a simple
model, we obtain the branching ratios $B(K_L \to  3 \gamma) \sim
3 \cdot 10^{-19}$, $B(K_S \to 3 \gamma) \sim 5 \cdot 10^{-22} $.
\end{abstract}

\newpage
The decays $K_L \to 2 \gamma $ and $K_S \to 2 \gamma $ have been observed with
branching ratios of $5.7 \cdot 10^{-4} $ and $2.4 \cdot 10^{-6} $,
respectively. What is the expected rate of the decays $K_{L,S} \to 3 \gamma $ ?
\par
\bigskip
First of all, one should observe that both $K_L \to 3 \gamma $ and $K_S \to
3 \gamma $ are possible without violating $CP $ invariance or any general
symmetry principle. Since the $3 \gamma $ system has $C = -1 $, the decay
$K_L \to 3 \gamma $ can proceed via the $C $--violating, $P $--violating
part of the $|\Delta S| = 1 $ nonleptonic weak interaction, while $K_S \to
3 \gamma $ proceeds via the $C $--conserving, $P $--conserving part.
Naively, one would imagine that these decays would occur at rates that are
roughly a factor $\alpha_{em} $ times the two--photon decay rates. \par
\bigskip
This naive expectation, however, disregards the constraints of gauge invariance
and Bose statistics. Gauge invariance dictates that in the decay $K^0 \to 3
\gamma $, no pair of photons can have angular momentum zero, since that would
correspond to a $0 \to 0 $ radiative transition, which is forbidden for a
real photon. Similarly, no pair of photons can have $J = 1 $, since that
conflicts with Bose statistics (Yang's theorem). It follows that the
decays $K_{L,S} \to 3 \gamma $ can only occur if each pair of photons in the
final state has at least two units of angular momentum. The matrix element
thus inevitably has a large number of angular momentum suppression factors.
Using a simple model, we show below that the decays $K_{L,S} \to 3 \gamma $
have rates that are {\bf 15 orders of magnitude lower} than the corresponding
rates of $K_{L,S} \to 2 \gamma $ ! \par
\bigskip
The model we employ is illustrated in Fig. 1. We assume that the $K_{L,S}
\to 3 \gamma $ transition is mediated by the decay $K_{L,S} \to \pi^0
\pi^0 \gamma $, with the two $\pi^0 $'s converting into two photons. The use
of this particular channel is motivated by the fact that the decays
$K_{L,S} \to \pi^0 \pi^0 \gamma $ are necessarily quadrupole transitions
(E2 and M2 respectively \cite{own}), so that the pion pair has $J = 2$, which
is the minimum angular momentum required for the photon pairs in $K_{L,S}
\to 3 \gamma $. Other intermediate states are undoubtedly possible
(including e.g. $K_{L,S} \to \pi^+ \pi^- \gamma $, with pions in a D--wave).
Our aim, however, is to expose the symmetry structure of the $K^0 \to 3
\gamma $ amplitude and to obtain a rough estimate of its magnitude. The
$\pi^0 \pi^0 \gamma $ intermediate state turns out to be adequate to this
purpose. \par
\bigskip
To determine the matrix element of $K_L \to 3 \gamma $ from the model shown
in Fig. 1, we require the amplitudes for $K_L \to \pi^0 \pi^0 \gamma $ and
$\pi^0 \pi^0 \to \gamma \gamma $, which we parameterise as follows:
\begin{eqnarray}
& & {\cal M} \left ( K_L(p_K) \to \pi^0(p_1) \; \pi^0(p_2) \;
\gamma(k_3,\epsilon_3) \right )
\;\;\;\;\;\;\;\;\;\;\;\;\;\;\;\;\;\;\;\;\;\;\;\;\;\;\;\;\;\;\;\;\;\;\;\;\;\;
\;\;\;\;\;\;\;\;\;\;\; \nonumber  \\
& = & {h_L \over M_K^5} \left [\epsilon_3 \cdot p_1 k_3 \cdot p_2 - \epsilon_3
\cdot p_2 k_3 \cdot p_1 \right ] \; k_3 \cdot (p_1 - p_2)
\end{eqnarray}
\begin{eqnarray}
& & {\cal M} \left ( \gamma (k_1,\epsilon_1) + \gamma (k_2, \epsilon_2)
\to \pi^0 (p_1) \; \pi^0(p_2) \right )  \nonumber \\
& = & {\tilde{G} s_{12} \over M_V^2} \cdot \left [ {p_1 \cdot k_1 \; p_1 \cdot
k_2 \over k_1 \cdot k_2} \; g_{\mu \nu} + p_{1 \mu} p_{1 \nu} - {p_1 \cdot k_1
\over k_1 \cdot k_2} \; k_{2 \mu} p_{1 \nu} - {p_1 \cdot k_2 \over k_1 \cdot
k_2} \; k_{1 \nu} p_{1 \mu} \right ] \epsilon_1^{\mu} \; \epsilon_2^{\nu}
\;\;\;\;\;\;
\end{eqnarray}
The structure in Eq. (1) is appropriate to an E2 transition, while that in
Eq. (2) is obtained using a vector meson exchange model for $\pi^0 \pi^0
\to \gamma \gamma $ (see, e.g., Ref. \cite{ko}), keeping only the leading
term in $s_{12}/M_V^2 $ ($s_{12} \equiv (k_1 + k_2)^2 $). (Numerically, the
constant $\tilde{G} $ is given by $\tilde{G} = \tilde{G}_{\rho} +
\tilde{G}_{\omega} $ with $\tilde{G}_{\rho} = {1\over 9} \;
g_{\omega \pi \gamma}^2, \; \tilde{G}_{\omega} = g_{\omega \pi \gamma}^2 $ and
$g_{\omega \pi \gamma} = 7.7 \cdot 10^{-4} \; \mbox{MeV}^{-1} $.) \par
\bigskip
We now employ unitarity to obtain the absorptive part of $K_L \to 3 \gamma $:
\begin{eqnarray}
& & Im \; {\cal M} \left [ K_L(p_K) \to \gamma(\epsilon_1, k_1) + \gamma
(\epsilon_2, k_2) + \gamma(\epsilon_3, k_3) \right ] \nonumber \\
& = & {1\over 2} \int {d^3p_1 \over 2 p_{1\, 0} (2\pi )^3} \; {d^3p_2 \over
2 p_{2\, 0} (2\pi )^3} \; (2\pi )^4 \delta^{(4)} (p_K - p_1 - p_2 - k_3)
\nonumber \\
& & \cdot \left \{ {h_L \over M_K^2} (\epsilon_3 \cdot p_1 \; k_3 \cdot p_2
- \epsilon_3 \cdot p_2 \; k_3 \cdot p_1) \; k_3 \cdot (p_1 - p_2) \right \}
\nonumber \\
& & \cdot {\tilde{G} s_{12} \over M_V^2} \; \epsilon_1^{\mu} \epsilon_2^{\nu}
\left \{ {p_1 \cdot k_1 \; p_2 \cdot k_2 \over k_1 \cdot k_2}\; g_{\mu \nu}
+ p_{1 \mu} p_{2 \nu} - {p_1 \cdot k_1 \over k_1 \cdot k_2}\; k_{2 \mu}
p_{1 \nu} - {p_1 \cdot k_2 \over k_1 \cdot k_2} \; k_{1 \nu} p_{1 \mu}
\right \} \nonumber \\
& & \cdot \Theta (s_{12} - 4 m_{\pi}^2) \nonumber \\
& & + \; \mbox{permutations of} \; (\epsilon_1, k_1), (\epsilon_2, k_2),
(\epsilon_3,k_3)
\end{eqnarray}
To evaluate this, we need to calculate integrals of the form
\begin{eqnarray}
K^{\mu \nu \rho \sigma} & = & \int {d^3p_1 \over 2 p_{1\, 0}} \;
{d^3p_2 \over 2 p_{2\, 0}} \; \delta^{(4)} (P - p_1 -p_2) f(p_1 \cdot p_2)\;
p_1^{\mu} p_1^{\nu} p_1^{\rho} p_1^{\sigma} \; , \nonumber \\
L^{\mu \nu \rho \sigma} & = & \int {d^3p_1 \over 2 p_{1\, 0}} \;
{d^3p_2 \over 2 p_{2\, 0}} \; \delta^{(4)} (P - p_1 -p_2) f(p_1 \cdot p_2)\;
p_1^{\mu} p_1^{\nu} p_1^{\rho} p_2^{\sigma}
\end{eqnarray}
These are given in the appendix. The resulting expression for ${\cal M}_{abs}
\equiv Im \; {\cal M} (K_L \to \gamma \gamma \gamma) $ is then squared, and
the polarizations of the photons summed over, using the symbolic computation
program FORM \cite{form}, with the result
\begin{eqnarray}
\sum_{pol} |{\cal M}_{abs}|^2 & = & V_{12}^2 F_{12}^2 \{s_{12} s_{23}^3 s_{13}
                                    +s_{12} s_{23} s_{13}^3 \} \nonumber \\
                              & + & V_{23}^2 F_{23}^2 \{s_{12}^3 s_{23} s_{13}
                                    +s_{12} s_{23} s_{13}^3 \} \nonumber \\
                              & + & V_{13}^2 F_{13}^2 \{s_{12}^3 s_{23} s_{13}
                                    +s_{12} s_{23}^3 s_{13} \} \nonumber \\
& - & 2 V_{12} V_{23} F_{12} F_{23} s_{12} s_{23} s_{13}^3 \nonumber \\
& - & 2 V_{12} V_{13} F_{12} F_{13} s_{12} s_{23}^3 s_{13} \nonumber \\
& - & 2 V_{13} V_{23} F_{13} F_{23} s_{12}^3 s_{23} s_{13}
\end{eqnarray}
where the functions $V_{ij} $ and $F_{ij} $ are defined by
\begin{eqnarray}
V_{ij} & = & {1\over 16 \pi} {h_L \over M_K^5} {\tilde{G} \over M_V^2}
{1 \over s_{ij}^2} \sqrt{ \lambda (s_{ij}, m_{\pi}^2, m_{\pi}^2) } \nonumber \\
F_{ij} & = & {1\over 5}\, s_{ij} \left \{ {1\over 8}\, s_{ij}^2 - {29 \over
24} \, s_{ij} m_{\pi}^2 + 2 m_{\pi}^4 \right \} \nonumber \\
\mbox{with} & & \lambda (x,y,z) = x^2 + y^2 + z^2 - 2xy - 2yz - 2xz
\end{eqnarray}
The following features of the above result (5) should be noted:
\begin{itemize}
\item[(i)] The Dalitz plot density, given by $\sum |{\cal M}_{abs}|^2 $, is
manifestly symmetric in $s_{12}, s_{23} $ and $s_{13} $, as required by
Bose statistics.
\item[(ii)] The density vanishes at the centre of the Dalitz plot ($s_{12}
= s_{23} = s_{13} $), i.e. for the configuration in which the photons have
equal energy.
\item[(iii)] Because of the overall factor $s_{12} s_{23} s_{13} $, the
density vanishes along the boundaries of the Dalitz plot, defined by
$s_{12} = 0, s_{23} = 0 $ and $s_{13} = 0 $. These correspond to the
configurations in which two of the three photons are collinear.
\item[(iv)] In the limit $m_{\pi} \to 0 $, the factors $V_{ij} F_{ij} $ become
proportional to $s_{ij}^2 $, and the density simplifies to
\begin{eqnarray}
\sum |{\cal M}_{abs}|^2 \stackrel{m_{\pi}= 0}{\sim} s_{12} s_{23} s_{13}
& & \left \{ s_{12}^2 \; (s_{12}^2 s_{23}^2 + s_{12}^2 s_{13}^2 - 2 s_{23}^2
s_{13}^2) \right. \nonumber \\
& & + s_{23}^2 \; (s_{23}^2 s_{12}^2 + s_{23}^2 s_{13}^2 - 2 s_{12}^2 s_{13}^2
) \nonumber \\
& & \left. + s_{13}^2 \; (s_{13}^2 s_{12}^2 + s_{13}^2 s_{23}^2 - 2 s_{12}^2
s_{23}^2 ) \right \}
\end{eqnarray}
\item[(v)] The result (7) has some resemblance to the matrix element for
$\pi^0 \to 3 \gamma $ obtained by Dicus \cite{dic}:
\newpage
\begin{eqnarray}
\sum |{\cal M}|^2\,|_{\mbox{Dicus}}  & \sim & s_{12} s_{13} s_{23} \;
\{ s_{12}^2 s_{13}^2 + s_{12}^2 s_{23}^2 + s_{23}^2 s_{13}^2 \nonumber \\
& & - s_{13} s_{23} s_{12}^2 - s_{13}^2 s_{23} s_{12} - s_{13} s_{23}^2
s_{12} \}
\end{eqnarray}
In concordance with Ref. \cite{dic}, we find that the $3 \gamma $ matrix
element contains a large number of momentum factors, which are ultimately
responsible for an enormous suppression of the decay rate. (The result given
by our model has two extra powers of $s_{ij} $ compared to the expression
in Eq. (8).)
\end{itemize}
Finally, we obtain the rate of $K_L \to 3 \gamma $ using
\begin{equation}
{d \Gamma \over ds_{12}\, ds_{23} } = {1\over (2 \pi )^3} { 1\over 32 M_K^3}
{1\over 3!} \sum |{\cal M}|^2
\end{equation}
and assuming that $\sum |{\cal M}|^2 $ is reasonably approximated by the
absorptive part given in Eq. (5). The parameter $h_L $ is determined by the
decay rate of $K_L \to \pi^0 \pi^0 \gamma $, for which we use the theoretical
estimate $B(K_L \to \pi^0 \pi^0 \gamma ) \sim 1 \cdot 10^{-8} $ obtained in
\cite{own}. The resulting branching ratio for $K_L \to 3 \gamma $ is
\begin{equation}
B(K_L \to 3 \gamma ) \sim 3 \cdot 10^{-19}
\end{equation}
The above considerations can be repeated for the decay $K_S \to 3 \gamma $, the
only difference being that the E2 matrix element given in Eq. (1) has to be
replaced by the M2 matrix element for $K_S \to \pi^0 \pi^0 \gamma $:
\begin{equation}
{\cal M} (K_S (p_K) \to \pi^0(p_1) \pi^0 (p_2) \gamma (\epsilon, k)
= {h_S \over M_K^5} \, (p_1 - p_2) \cdot k \, \epsilon_{\mu \nu \rho \sigma}
\epsilon^{\mu} k^{\nu} p_1^{\rho} p_2^{\sigma}
\end{equation}
The Dalitz plot density turns out to have exactly the same functional
dependence on $s_{12}, s_{23} $ and $s_{13} $ as in Eq. (5). The branching
ratio is estimated to be
\begin{equation}
B(K_S \to 3 \gamma ) \sim 5 \cdot 10^{-22}
\end{equation}
The exceedingly low rates given by Eqs. (10) and (12) imply that the three
photon decay of the neutral $K $ meson cannot be a significant background to
decays of the type $K^0 \to 2 \pi^0 \to 4 \gamma $, in which one photon is
undetected. \par
\bigskip
\bigskip

{\bf \large Acknowledgement} \par
\bigskip
This work was initiated during a visit by one of us (L.M.S.) to the University
of Melbourne. The hospitality of the School of Physics is gratefully
acknowledged. The research has been supported by the German Ministry of
Research and Technology.

\newpage
{\large \bf Appendix} \par

\bigskip
\noindent
We give here the integrals defined in Eq. (4), for the general case $p_1^2
= m_1^2, \; p_2^2 = m_2^2 $.
\begin{eqnarray}
K^{\mu \nu \rho \sigma} & = & \int {d^3 p_1 \over 2 p_{1 \, 0} }
{d^3 p_2 \over 2 p_{2 \, 0} } \delta^{(4)} \; (P - p_1 -p_2) \;
f(p_1\cdot p_2) \; p_1^{\mu} p_1^{\nu} p_1^{\rho} p_1^{\sigma} \nonumber
\\
& = & {\pi \over 2} {1\over s^3} \sqrt{\lambda (s, m_1^2, m_2^2)}
f \left [ {1\over 2} (s - m_1^2 -m_2^2) \right ] \; \cdot {1\over 5} \nonumber
\\
& & \cdot \left \{ {1\over s^2} \left [ (s + m_1^2 - m_2^2)^4 - 3 s m_1^2
(s+ m_1^2 - m_2^2)^2 + s^2 m_1^4 \right ] P^{\mu} P^{\nu} P^{\rho} P^{\sigma}
\right. \nonumber \\
& & - \left [ {1\over 8} {1\over s} (s + m_1^2 - m_2^2)^4 - {7 \over 12} m_1^2
(s + m_1^2 - m_2^2)^2 + {1\over 3} s m_1^4 \right ] \nonumber \\
& & \cdot \left (P^{\mu} P^{\nu} g^{\rho \sigma} + P^{\nu} P^{\rho}
g^{\mu \sigma} + P^{\mu} P^{\rho} g^{\nu \sigma} + P^{\nu} P^{\sigma}
g^{\mu \rho} + P^{\rho} P^{\sigma} g^{\mu \nu} + P^{\mu} P^{\sigma}
g^{\nu \rho} \right ) \nonumber \\
& & + {1\over 3} \left [ {1\over 16} (s + m_1^2 - m_2^2)^4 - {1\over 2} s
m_1^2 (s+ m_1^2-m_2^2)^2 + s^2 m_1^4 \right ] \nonumber \\
& &  \cdot \left ( g^{\mu \nu} g^{\rho \sigma} + g^{\nu \rho}
g^{\mu \sigma} + g^{\mu \rho} g^{\nu \sigma } \right ) {\Large \}} \nonumber
\end{eqnarray}

\begin{eqnarray}
L^{\mu \nu \rho \sigma} & = & \int {d^3 p_1 \over 2 p_{1 \, 0} }
{d^3 p_2 \over 2 p_{2 \, 0} } \delta^{(4)} \; (P - p_1 - p_2) \;
f(p_1 \cdot p_2) \; p_1^{\mu } p_1^{\nu } p_1^{\rho } p_2^{\sigma} \nonumber
\\
& = & {\pi \over 2} {1 \over s^3} \sqrt{\lambda (s, m_1^2, m_2^2)}
f \left [ {1\over 2} (s - m_1^2 - m_2^2) \right ] \; \cdot {1 \over 5}
\nonumber \\
& & \cdot \left \{ {1\over s^2} \left [ (s + m_1^2 - m_2^2)^3
(s - m_1^2 + m_2^2)^2 - {3\over 4} s \left \{ (s+ m_1^2 - m_2^2)^2 \right.
\right. \right. \nonumber \\
& & \;\;\;\;\;\;\;\; \left. \cdot (s - m_1^2 -m_2^2) + m_1^2
(s + m_1^2 -m_2^2) (s - m_1^2 + m_2^2) \right \} \nonumber \\
& & \;\;\;\;\;\;\;\; \left. + {1\over 2} s^2 m_1^2 (s - m_1^2 - m_2^2) \right ]
\; P^{\mu} P^{\nu} P^{\rho} P^{\sigma} \nonumber \\
& & \;\;\; - \left [ {1\over 8} {1\over s} (s + m_1^2 - m_2^2)^3 (s - m_1^2
+ m_2^2) - {7\over 48} \left \{ (s + m_1^2 - m_2^2)^2 \right. \right.
\nonumber \\
& & \;\;\;\;\;\;\; \left. \cdot (s - m_1^2 -m_2^2) + m_1^2 (s + m_1^2 - m_2^2)
(s - m_1^2 + m_2^2) \right \} \nonumber \\
& & \;\;\;\;\;\;\; \left. + {1\over 6} s m_1^2 (s - m_1^2 - m_2^2) \right ]
\nonumber \\
& & \cdot (P^{\mu} P^{\nu} g^{\rho \sigma} + P^{\nu} P^{\rho} g^{\mu \sigma}
+ P^{\mu} P^{\rho} g^{\nu \sigma} + P^{\nu} P^{\sigma} g^{\mu \rho}
+ P^{\rho} P^{\nu} g^{\mu \nu} + P^{\mu} P^{\sigma} g^{\nu \rho} )
\nonumber \\
& & + {1\over 3} \left [ {1\over 16} ( s+ m_1^2 - m_2^2)^3 (s - m_1^2 + m_2^2)
- {1\over 8} s \left \{ (s + m_1^2 - m_2^2)^2 \right. \right. \nonumber \\
& & \;\;\;\;\;\;\; \left. \cdot (s - m_1^2 - m_2^2) + m_1^2 (s + m_1^2
- m_2^2) (s - m_1^2 + m_2^2) \right \} \nonumber \\
& & \;\;\;\;\;\;\; \left. \left. + {1\over 2} s^2 m_1^2 (s - m_1^2 - m_2^2)
\right ] \; \cdot (g^{\mu \nu} g^{\rho \sigma} + g^{\nu \rho} g^{\mu \sigma}
+ g^{\mu \rho} g^{\nu \sigma} ) \right \} \nonumber
\end{eqnarray}


\newpage
\begin{thebibliography}{99}
%
\bibitem{own} P. Heiliger and L.M. Sehgal, Phys. Lett. B 307 (1993) 182.
\bibitem{ko} P. Ko, Phys. Rev. D 41 (1990) 1531.
\bibitem{form} J.A.M. Vermaseren, Symobolic Manipulation with FORM,
               CAN (1991).
\bibitem{dic} D.A. Dicus, Phys. Rev. D 12 (1975) 2133.
\end{thebibliography}
\end{document}